# Intermodulation Distortion of *Actuated* MEMS Capacitive Switches

Xi Luo[1], Yaqing Ning[1], David Molinero[1], Cristiano Palego[1], James C. M. Hwang[1], and Charles L. Goldsmith[2]

[1]Lehigh University, Bethlehem, Pennsylvania, 18015, USA
[2]MEMtronics Corporation, Richardson, Texas, 75081, USA

*Abstract* — For the first time, intermodulation distortion of micro-electromechanical capacitive switches in the *actuated* state was analyzed both theoretically and experimentally. The distortion, although higher than that of switches in the *suspended* state, was found to decrease with increasing bias voltage but to depend weakly on modulation frequencies between 55 kHz and 1.1 MHz. This dependence could be explained by the orders-of-magnitude increase of the spring constant when the switches were actuated. Additionally, the analysis suggested that increasing the spring constant and decreasing the contact roughness could improve the linearity of actuated switches. These results are critical to micro-electromechanical capacitive switches used in tuners, filters, phase shifters, etc. where the linearity of both suspended and actuated states are critical.

## I. INTRODUCTION

Capacitive switches based on micro-electromechanical systems (MEMS) are known [1], [2] to be extremely linear with the third-order intercept as high as 80 dBm, so long as the modulation frequency is too high for the switches to respond mechanically. (Typically, the mechanical resonance frequency $f_0$ ~ 100 kHz.) However, these observations were mainly based on switches in the *suspended* state and under small-signal conditions. Under large-signal conditions, higher-order intermodulation could be generated to degrade linearity [3]. The linearity could degrade further when the switches were biased, which made the mechanical response highly nonlinear [4]. This paper reports, for the first time, the linearity of MEMS capacitive switches when the bias voltage is increased above the pull-in voltage so that the switches are *actuated*. If the switches are used merely as on/off switches in the shunt configuration, then only the linearity of the *suspended* state is critical. However, in other configurations such as in tuners, filters, phase shifters, etc., the linearity of both *suspended* and *actuated* states are critical.

## II. CAPACITANCE-VOLTAGE CHARACTERISTICS

The present MEMS capacitive shunt switches are based on a movable aluminum membrane 0.3-μm thick and 120-μm wide, which is fixed at both ends of its length $\ell$ = 310 μm [5]. Without any bias, the membrane is on average suspended $g$ = 2.2 μm above a stationary gold electrode, which is coated with 0.25-μm-thick SiO$_2$ to prevent DC short when the membrane is pulled in by DC bias to contact the stationary electrode. Typically, pull-in voltage $V_P$ ~ 25 V and release voltage $V_R$ ~ 11 V.

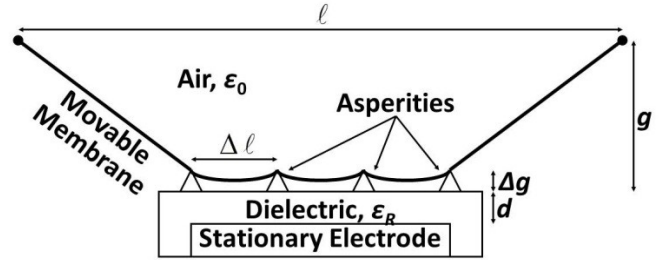

Fig. 1. Greatly simplified contact scenario in which the contact is mainly at asperities and bias higher than the pull-in voltage makes the contact more intimate and, hence, the contact capacitance higher.

When a switch is actuated, the exact nature of the contact is complicated by the details of the contact surfaces. Fig. 1 shows a greatly simplified case in which the contact is mainly at asperities and biases higher than $V_P$ makes the contact more intimate and, hence, the capacitance across the membrane and the stationary electrode higher. Meanwhile, the membrane is effectively fixed at the asperities and the spring constant is effectively that of each section between the asperities. Since the spring constant $k$ ~ $1/\ell^3$, the asperity height $\Delta g$ ~ 0.1 μm, and the asperity spacing $\Delta\ell$ ~ 1 μm [6], $k'$, the spring constant after pull in, can be a million times or higher than $k$. This can, in turn, increase the mechanical resonance frequency beyond the microwave range. The above-described increases in contact capacitance and spring constant can explain the bias and frequency dependence of the linearity of actuated switches as described in the following.

Fig. 2 shows the measured capacitance-voltage characteristic of a typical MEMS capacitive switch. With the bias voltage $V_{DC} = V_P$, the switch capacitance $C$ increases abruptly. With $V_{DC} > V_P$, the capacitance continues to increase gradually with increasing $V_{DC}$. Thus, $C$ in the actuated state can be empirically expressed in constant and voltage-dependent terms as

$$C = \varepsilon_0 \varepsilon_R \frac{A}{d}\left(1 - \varepsilon_R \frac{\Delta g}{d} + \frac{V_{DC}}{V_0}\right) \qquad (1)$$

where $\varepsilon_0$ is vacuum permittivity, $A$ is the contact area, $d$ is the dielectric thickness, $\varepsilon_R$ is the dielectric constant, and $V_0$ is a measure of voltage dependence. For the present switches, $A = 120\ \mu m \times 80\ \mu m$, $d = 0.25\ \mu m$, and $\varepsilon_R = 5$. The first term, $\varepsilon_0\varepsilon_R A/d$, is the capacitance of an ideal contact. The second term, $\varepsilon_R \Delta g/d$, corrects for surface roughness and contamination. For the present switches with aluminum membranes, $\Delta g \sim 0.3\ \mu m$ [7]. However, for more recent switches with molybdenum membranes and an associated higher temperature and cleaner release process [8], $\Delta g \sim 0.1\ \mu m$ which reflects the true surface roughness. Fig. 2 shows also that the measured voltage dependence of the switch capacitance can be fitted with $V_0 = 517$ V.

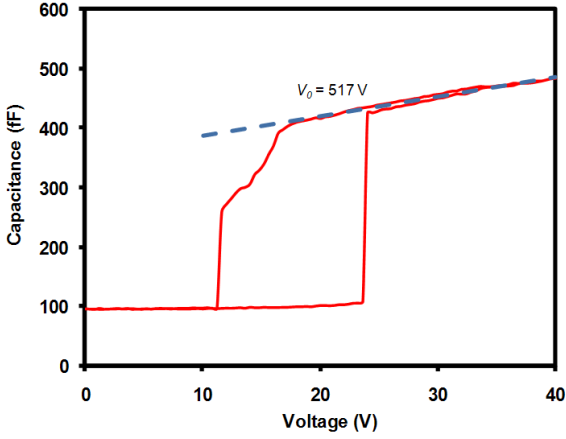

Fig. 2. Measured capacitance-voltage characteristic of a typical MEMS capacitive switch.

### III. MICROWAVE INTERMODULATION MEASUREMENT

Fig. 3 illustrates the microwave intermodulation measurement setup [5]. Unpackaged switches are tested on-wafer. DC bias $V_{DC}$, modulation signal $V_{LF}$ ($f_{LF}$ = 55 kHz to 1.1 MHz), and microwave signal $V_{RF}$ ($f_{RF}$ = 6 GHz) are combined at the input of each switch, so that

$$V_{IN} = V_{DC} + V_{LF}\sin(\omega_{LF}t) + V_{RF}\sin(\omega_{RF}t) \quad (2)$$

where $t$ is time. The microwave power is limited to 0.1 W, which is a small fraction of the nominal power-handling capacity of the switch. With the switch actuated, most of the microwave power is reflected, which is sampled by a spectrum analyzer. The measured third-order intermodulation product (IMD3) should equal to that of a two-tone test with $V_1 = V_2 = V_{LF}/\sqrt{2}$, $f_1 = f_{RF}$ and $f_1 - f_2 = f_{LF}$.

### IV. INTERMODULATION DISTORTION ANALYSIS

The reflection ratio from an actuated switch is

$$S_{11} = \frac{-j\omega_{RF}CZ_0}{2+j\omega_{RF}CZ_0} = -\frac{2j\omega_{RF}CZ_0 + (\omega_{RF}CZ_0)^2}{4+(\omega_{RF}CZ_0)^2} \quad (3)$$

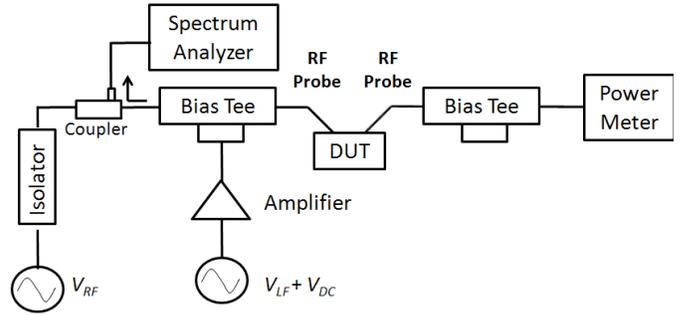

Fig. 3. Microwave intermodulation measurement setup.

where $j = \sqrt{-1}$, $\omega_{RF} = 2\pi f_{RF}$, and $Z_0$ is the system impedance. The phase of the reflected signal is

$$\phi_0 + \Delta\phi = \tan^{-1}\frac{2}{\omega_{RF}(C_0 + \Delta C)Z_0}$$

$$\approx \frac{2}{\omega_{RF}(C_0 + \Delta C)Z_0} \quad (4)$$

$$\approx \frac{2}{\omega_{RF}C_0 Z_0}\left(1 - \frac{\Delta C}{C_0}\right)$$

where

$$C_0 = \varepsilon_0\varepsilon_R \frac{A}{d}\left(1 - \varepsilon_R \frac{\Delta g}{d}\right) \quad (5)$$

and

$$\frac{\Delta C}{C_0} = \frac{V_{IN}^{rms}}{V_0\left(1-\varepsilon_R\frac{\Delta g}{d}\right)}$$

$$= \frac{\sqrt{V_{DC}^2 + \frac{V_{LF}^2}{2} + \frac{V_{RF}^2}{2} + 2V_{DC}V_{LF}\sin(\omega_{LF}t) - \frac{V_{LF}^2}{2}\cos(2\omega_{LF}t)}}{V_0\left(1-\varepsilon_R\frac{\Delta g}{d}\right)}. \quad (6)$$

Therefore,

$$\Delta\phi \approx -\phi\frac{\Delta C}{C_0} \approx -\frac{2\Delta C}{\omega_{RF}C_0^2 Z_0} \quad (7)$$

and IMD3 at $\omega_{RF} \pm 2\omega_{LF}$ is

$$P_{IMD3} = \left(\frac{V_{LF}^2}{4\omega_{RF}C_0 Z_0 V_0 V_{DC}(1-\varepsilon_R\frac{\Delta g}{d})}\right)^2. \quad (8)$$

From (8), linearity improves with increasing $\omega_{RF}$, $C_0$, $V_0$, and $V_{DC}$. On the other hand, linearity degrades with increasing $\Delta g$.

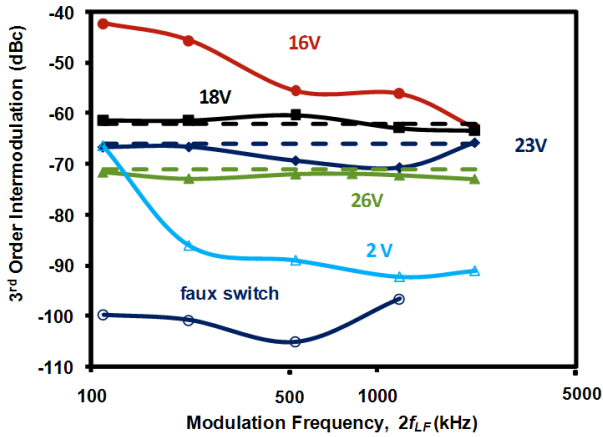

Fig. 4. Measured IMD3 of a switch in suspended ($V_{DC}$ = 2 V), actuated ($V_{DC}$ = 26 V), and intermediate ($V_{DC}$ = 18 V) states, respectively, vs. that of a faux switch with its contact permanently closed. $P_{RF}$ = 20 dBm. $P_{LF}$ = 23 dBm.

## V. RESULTS AND DISCUSSION

Fig. 4 shows the measured IMD3 for a switch in suspended ($V_{DC} < V_R$), actuated ($V_{DC} > V_P$), and intermediate ($V_R < V_{DC} < V_P$) states, respectively. It can be seen that, with $V_{DC}$ = 2 V, the membrane is suspended and IMD3 exhibits the familiar low-pass behavior and, at $f_{LF} \gg f_0$, approaches that of a faux switch with its contact permanently closed thereby resembling an ideal contact [9]. With $V_{DC}$ = 26 V, the membrane is pulled-in and IMD3 degrades by approximately 20 dB and exhibits little frequency dependence. When $V_{DC}$ is then reduced to 23 V, the membrane remains pulled in, but is not held down as tightly as the case of 26 V. Therefore, IMD3 degrades by approximately another 10 dB. Further degradation occurs at 18 V. Finally, low-pass behavior is observed at 16 V indicating that the membrane is partially released. Fig. 4 shows also that IMD3 can be fitted with (8) between 18 V and 26 V.

The present analysis is based on non-ideal contact to a rough surface. The result suggests that decreasing the contact roughness and/or increasing the membrane stiffness should improve the linearity in general. The present analysis is equally applicable to contacts with purposely fabricated standoffs (bumpers) to minimize dielectric charging [10]. In this case, the height and spacing of the standoffs need to be carefully traded off for the optimum combination of performance and reliability.

## VI. CONCLUSION

For widely spaced tones and as simple on/off shunt switches, the linearity of MEMS capacitive switches is mainly limited by the *suspended* state which is indeed excellent. As tuners, filters, phase shifters, etc., the linearity is mainly limited by the *actuated* state, which can be worse by 20 dB or more. For wideband modulations, the linearity is limited by the *suspended* state at low modulation frequencies but by the *actuated* state at high modulation frequencies. In this case, both *suspended* and *actuated* states are critical.

ACKNOWLEDGMENT

This work was sponsored in part by a fellowship from the Automatic RF Techniques Group, as well as by the Defense Advanced Research Projects Agency (DARPA) N/MEMS S&T Fundamentals program under grant no. N66001-10-4006 issued by the Space and Naval Warfare Systems Center Pacific (SPAWAR).